\documentstyle[12pt,aaspp4,astrobib,epsfig,here]{article}

\newcommand{\scrm}[1]{\mbox{\scriptsize\rm #1}}
\def\ltsima{$\; \buildrel < \over \sim \;$}
\def\simlt{\lower.5ex\hbox{\ltsima}} 
\def\gtsima{$\; \buildrel > \over \sim \;$}
\def\simgt{\lower.5ex\hbox{\gtsima}} 

\def\K{\hbox{$\rm K$}}
\def\J{\hbox{$\rm J$}}
\def\V606{\hbox{$\rm V_{606}$}}
\def\I814{\hbox{$\rm I_{814}$}}
\def\I{\hbox{\rm I}}
\def\V{\hbox{$\rm V$}}
\def\H{\hbox{$\rm H$}}
\def\K{\hbox{$\rm K$}}

\def\AV{\hbox{\rm A}_V}
\def\EBV{$E(B\!-\!V)$}

\def\Msun{\:{M_{\odot}}}

\newcommand{\etal} {{\it et~al.\ }}
\begin{document}

\title{The Fading Optical Counterpart of GRB~970228, Six Months 
and One Year Later}

\author{Andrew S. Fruchter$^1$,
Elena Pian$^2$,
Stephen E. Thorsett$^3$,
Louis E. Bergeron$^1$,
Rosa A. Gonz\'alez$^1$,
Mark Metzger$^{4}$,
Paul Goudfrooij$^{1,5}$,
Kailash C. Sahu$^1$,
Henry Ferguson$^{1}$,
Mario Livio$^1$,
Max Mutchler$^1$,
Larry Petro$^1$,
Filippo Frontera$^{2,6}$,
Titus Galama$^7$,
Paul Groot$^7$,
Richard Hook$^8$,
Chryssa Kouveliotou$^9$,
Duccio Macchetto$^{1,5}$,
Jan van Paradijs$^{7}$,
Eliana Palazzi$^{2}$,
Holger Pedersen$^{10}$,
William Sparks$^1$, 
Marco Tavani$^{11,12}$}

\affil{$^{1}$Space Telescope Science Institute, 3700 San Martin 
Drive, Baltimore, MD 21218, USA\\
$^{2}$Istituto di Tecnologie e Studio delle Radiazioni 
Extraterrestri, C.N.R., Via Gobetti 101, I-40129 Bologna, Italy \\
$^{3}$Joseph Henry Laboratories and Dept.\ of Physics, Princeton 
University, Princeton, NJ 08544, USA\\
$^{4}$Department of Astronomy, Caltech, MS 105-24, Pasadena, CA 91125\\
$^{5}$Affiliated to the Astrophysics Division, Space Science Department, 
European Space Agency\\
$^{6}$Dip. Fisica, Universit\`a di Ferrara, Via Paradiso 
12, I-44100 Ferrara, Italy\\
$^{7}$Astronomical Institute ``Anton Pannekoek'', University
of Amsterdam, Kruislaan 403, 1098 SJ Amsterdam, The Netherlands\\
$^8$Space Telescope European Coordinating Facility, 
Karl-Schwarzschild-Str. 2, D-85748 Garching, 
Germany\\
$^9$NASA Marshall Space Flight Center, ES-84, Huntsville, AL 35812, USA\\
$^{10}$Copenhagen University Observatory, Juliane Maries Vej 30, D-2100, 
Copenhagen \"{A}, Denmark \\
$^{11}$Columbia Astrophysics Laboratory, Columbia 
University, New York, NY 10027, USA \\
$^{12}$Istituto di Fisica Cosmica e Tecnologie Relative,
C.N.R., Via Bassini 15, I-20133 Milano, Italy}

\begin{abstract}
We report on observations of the fading optical counterpart of the
gamma-ray burst GRB~970228, made  with the Hubble Space Telescope
and the Keck I telescope.
The GRB was observed approximately six months after outburst,
on  4~September~1997, using the HST/STIS
CCD, and approximately one year after
outburst, on 24~February~1998, using HST/NICMOS, and 
on 4 April 1998 using the NIRC on Keck.
The unresolved counterpart
is detected by STIS at $V=28.0\pm0.25$, consistent with a continued power-law
decline with exponent $-1.10 \pm 0.10$.  
The counterpart is located
within, but near the edge of, a faint extended source with 
diameter $\sim 0\farcs8$ and integrated magnitude
$V=25.8\pm0.25$. A reanalysis of HST and  NTT
observations performed shortly after the burst 
shows no evidence of proper motion of the point
source or fading of the extended emission.  
Although the optical transient (OT)
is not detected in the NICMOS images ($\H \ge 25.3$),  
the extended source
is visible and has a total magnitude $\H=23.3 \pm 0.1$. 
The Keck observations
find $\K = 22.8 \pm 0.3$.    
Comparison with 
observations obtained shortly after outburst suggests
that the nebular luminosity has also been stable in the infrared.

We find that 
several distinct and independent means of deriving the
foreground extinction in the direction of GRB~970228 all
agree with $A_V = 0.75 \pm 0.2$. 
After adjusting for this Galactic extinction,
we find that the size of the observed extended emission
is consistent with that of
galaxies of comparable magnitude found in the Hubble Deep Field (HDF) and
other deep HST images.  Only 2\% of the sky is covered
by galaxies of similar  or greater surface brightness.
We therefore conclude that the extended source observed about GRB~970228
is almost certainly its host galaxy.  
Additionally, we find that independent of
assumed redshift, the
host is significantly bluer than typical nearby blue dwarf irregulars.
With the caveat that the presently available infrared observations
of the HDF are only fully complete to a limit about one-half magnitude
brighter than the host, we find
that {\it the extinction-corrected $\V - \H$ and $\V - \K$ colors 
of the host 
are as blue as any galaxy of comparable or
brighter magnitude in the HDF}.
Taken in concert with recent observations of GRB~970508, GRB~971214, and
GRB~980703 
our
work suggests
that all four GRBs with spectroscopic identification or 
deep multicolor broad-band imaging of the host
lie in rapidly star-forming galaxies.     

\end{abstract}

\section*{Introduction}
The field of gamma-ray burst (GRB) astronomy was transformed in
early 1997, when sub-arcminute localization of the burst GRB~970228 by
the gamma- and X-ray instruments on the BeppoSAX satellite allowed the
first firm optical identification of a fading GRB counterpart
\cite{vgg+97}.  Within months, a second optical counterpart was found
\cite{bond6654,dmk+97}, for GRB~970508, allowing the first
spectroscopic limit on the distance to a GRB, $z\ge0.835$
\cite{mdk+97}. There are now nearly ten bursts with identified optical 
counterparts, including three with measured \nocite{kfwe+98,gvgs+98,hf98} 
redshifts.\footnote{The GRB~980425 has been associated with SN1998bw.
If this association is correct, then this source lies
in a nearby spiral galaxy with redshift
of $z = 0.008$ (Galama \etal 1998).  
Its X-ray and optical luminosity would be far below that
of the other GRBs with spectroscopic identitication, and it most likely
represents a different class of objects (Kulkarni \etal 1998, Hogg
and Fruchter 1998) from
the other GRBs discussed in this work.} 

The fading counterpart of GRB~970228 remains one of the best studied of
this new class of astronomical objects, but the observations have
been confusing.  Early HST imagery suggested the presence of a
nebulosity centered $\sim 0\farcs3$~arcsecond from the point-like fading
transient source \cite{slp+97}.  Tentative evidence was presented that
the nebular emission was fading \cite{mcb6676}, and a proper motion of
550~mas~yr$^{-1}$ was reported for the fading point source
\cite{cmtb97b}, though the measurement was disputed \cite{slpb+97}.
Either a fading nebula or a measurable proper motion would ineluctably
lead to the conclusion that GRB~970228 was a Galactic event, unlike
GRB~970508 and the other bursts with measured redshifts.

To help resolve the situation, and to further characterize the
broadband spectral properties of the GRB counterpart, we have
reobserved GRB~970228 with Keck and HST.  Although the earlier HST
observations of GRB~970228 employed WFPC2, we have availed ourselves of
the newly installed STIS CCD and NICMOS camera 2.  The excellent
throughputs, broad bandpasses and wide spectral coverage of these
instruments, combined with the long time baseline since the gamma-ray
burst, provide us with a superb opportunity to study the nature of the
source and its environment.

\section*{Observations and Image Analysis}

\subsection*{The STIS Images}
The field of GRB~970228 was imaged during two HST orbits
on 4 September 1997
from 15:50:33 to 18:22:41 UT, using the
STIS CCD in Clear Aperture (50CCD) mode.  Two exposures
of 575s each were taken at each of four dither positions
for a total exposure time of 4600s.  The exposures were dithered
to allow removal of hot pixels and to obtain
the highest possible resolution.  The images were bias and
dark subtracted,
and flat-fielded using the STIS
pipeline.  The final image was created and cleaned of cosmic rays
and hot pixels
using the variable pixel linear reconstruction 
algorithm (a.k.a. Drizzle) developed for the
Hubble Deep Field (HDF) \cite{wms96,fh97}.

\begin{figure}
\centerline{\epsfig{file=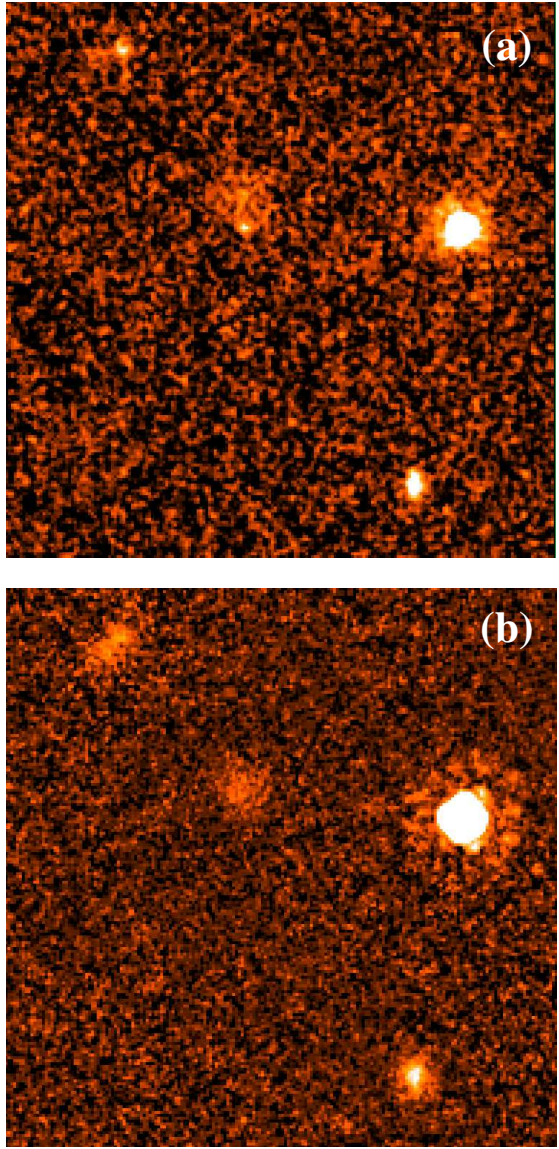,height=6.0in}}
\vspace{10pt}
\caption{The STIS, (a), and NICMOS, (b), drizzled
images of GRB 970228.  In both images,
North is up; East
is to the left.  The images are $7\farcs2$ on a side, and  
the pixel size is one half that of the original NIC-2 pixels,
or about $0\farcs0378$ per pixel.
The host galaxy is the roughly circular
nebulosity about $1\farcs5$ NE of the center of the image.
The OT can be seen in the STIS image as a faint point
source near
the southern edge of the host galaxy.  
In the NICMOS image, taken approximately
six months after the STIS image and one year after outburst, the OT is
no longer visible.}
\label{f:both}
\end{figure}

The photometric calibration of the images was performed using the synthetic
photometry package {\tt synphot} in IRAF/STSDAS,  although total throughput
was renormalized by approximately 12\% to agree with the on-orbit 
recalibration of the STIS CCD by
Landsman (1997)\nocite{land97}.
This adjustment
agrees well both with estimates derived by
comparison of the stellar magnitudes
in the WFPC2 F606W image of the GRB~970228 field with
those of the STIS image, and with an independent
 calibration of the STIS CCD
performed by Sahu \etal (1997)\nocite{slp+97}.  

The STIS CCD in clear aperture mode has a broad bandpass, with
a significant response from
200 to 900 nm that peaks near 600nm.  As a result,
STIS instrumental magnitudes are most accurately
translated into the standard filter set by quoting
the result as a $\V$ magnitude, but in any case
knowledge of an object's intrinsic spectrum is required 
for an accurate conversion.   Here
 we are fortunate to have an earlier
image of the field in both V(F606W) and I(F814W) with the
WFPC2.  We therefore can use colors determined
in these earlier observations to interpret the
STIS data.  However, to the extent that the WFPC2 colors are
in error (either due to low signal-to-noise ratio or an
actual change in the color of the object over time)
our derived magnitudes from the STIS data will be
biased by about 0.5 magnitudes for each magnitude
of error in $\V - \I$ \cite{ssk+98}.

The magnitude of the OT was determined
from the drizzled image via aperture photometry.  The flux
in an aperture of
radius four drizzled pixels, or $0\farcs1$, was determined, and
our best estimate of the surrounding nebular background was
subtracted.  An aperture correction of $0.50$ magnitudes
was derived for this aperture using the bright star
visible in Figure 1 to the west of the nebula.
We find that the OT has a count rate of $0.206 \pm 0.02$ counts per
second for a derived magnitude of
$\V = 28.0 \pm 0.25$, where the final error includes 
the uncertainty of
the conversion to the standard photometric bandpass.   

We determined the magnitude of the nebula
by summing all pixels in a region of approximately
$1.4$ sq. arcsec. surrounding the object.   The flux of the
point source was then subtracted
from the sum.   We find a flux of $1.36 \pm 0.1$ counts per second
and a magnitude of $\V = 25.8 \pm 0.25$ for
the nebula.  Again, the final error is dominated by 
the uncertainty of the $\V - \I$ color of the object, 
for which we have only the 
previous, comparatively noisy,
WFPC2 observations of
the field \cite{slp+97}.

However, the WFPC2 observations provide
us with a baseline of 162 days, and thus
an excellent opportunity to investigate
the variability and claimed 
proper motion of the OT, as well as to determine
whether the apparent magnitude of the extended emission
has remained constant.


\subsection*{The WFPC2 Images}

In order to search for proper motion of the OT
and possible changes in the apparent brightness of
the extended source, we have reexamined the WFPC2
images obtained in March and April of 1997 \cite{slp+97}.
The WFPC2 images 
were processed using the standard WFPC2 data pipeline,
and cosmic rays removed using the standard STSDAS task "crrej".
The centroids of the four reference stars used as positional
anchors by Sahu \etal (1997b) 
were redetermined using 2-dimensional
Gaussian fits to both the WFPC2 and STIS images. 
The positional accuracy for the reference stars is 
approximately 3 mas in each coordinate at each epoch, 
while for the relatively faint GRB it is $\sim$10 mas.  
The measured pixel-coordinates were corrected for the cameras' geometric 
distortion using the Gilmozzi \etal\ (1995) \nocite{gsk95}
solutions  for the WFPC2 images,
and a solution derived by Malamuth and Bowers  (1997) \nocite{mb97}
for the STIS images.  
The STIS pixel-coordinates were then transformed to the corresponding
WFPC2 pixel-coordinates, taking into account the rotation, translation and the 
image scale change while
 assuming zero mean motion of the four reference stars.
 This  procedure was performed on 
the WFPC2 V and I-band images separately.
The positions of the four stars in the WFPC2 images
agree with their positions in the STIS images  to within the expected 
uncertainties of 3 mas in both colors, 
which shows that the transformations between the two images have
been done correctly.  
Averaging together all of the data
we find that any motion of the GRB between the two epochs
is less than $16$~mas.
This corresponds to a motion of less than 36~mas
per year. This is a factor of $\sim 15$ less
than the value claimed by Caraveo \etal
(1997), and improves the upper limit on the proper motion
reported by Sahu \etal (1997b) by a factor of six.

To check on the previous WFPC2
photometry of the OT and extended emission,
the point
source magnitude was determined by using 
circular apertures of radii 1 and 3 pixels in the WFPC2 images.
The values obtained were then adjusted
by applying the aperture
corrections 
and in the case of F606W, the color term for transformation
to Johnson V, derived by Holtzman \etal (1995)\nocite{hbc+95}. 
The nebular magnitude was determined from the WFPC2 images
by taking the sum
of all counts above sky in a box approximately $1\farcs5 \times 1\farcs0$, and
subtracting the counts (estimated as above) attributable to
the point source.    The position of this box was determined
by the position of the nebula in the STIS image.  It is, however,
somewhat larger than the observed nebula in all directions.
Averaging together the two WFPC2 observations, we obtain magnitudes
for the extended emission of
$\I = 24.4 \pm 0.2$
and $\V = 25.8 \pm 0.2$.   This visual magnitude is
remarkably consistent with the STIS measurement.

\subsection*{The NICMOS Images}

On 24 February 1998, we obtained a total integration time 
of $\sim ~10,000$s, over four orbits, on the field of GRB~970228 
using the F160W (H) filter of NICMOS Camera 2.
A similar set of exposures in the F110W (J) filter, 
scheduled for the previous day,
were lost when a cosmic ray event in the the NICMOS electronics
 caused the camera to be reset during the
first of the four orbits. 
However, as the F160W band on NICMOS is extremely sensitive
and provides greater leverage in wavelength than
the F110W filter when
compared with WFPC2 and STIS imaging, the observations in hand
remain immensely useful.

Each of the four orbits was subdivided into a single $512$s and two
$1024$s exposures, with two dither positions performed each orbit.
All exposures were taken using multiple initial and final reads, or
MIF sequences.
In all, eight dither positions were placed roughly along a $\sim 1"$ diagonal 
line.  The images were processed using a variant of the standard
NICMOS pipeline.   
``Superdark" reference files were produced for the MIF512 and MIF1024
sequences using a sigma-clipped average (3 sigma) of 20 MULTIACCUM 
on-orbit darks in each sequence, taken in November 1997. 
First an estimate of the thermal background, the pedestal (a constant offset
in each of the four chip quadrants) and  shading (effectively a
ramp in bias across each quadrant) were subtracted from the
MULTIACCUM sequences.   Then flat-fielding, cosmic-ray rejection
and conversion to count-rates were performed.  When the final
images were combined using Drizzle,  noticeable streaking along the
diagonal of the dither, as well as large scale noise remained, even
after further image defects and residual cosmic-rays were  removed
using the cosmic-ray rejection scheme described in Fruchter and
Hook (1997).    

We therefore created ``sky'' images using a technique very similar to
that employed by ground-based infrared astronomers.   First, the
median of all MIF1024 and MIF512 images were separately obtained while
masking out the astrophysical objects detected in the drizzled image
previously created.  These images revealed significant structure in
the sky that differed substantially between the MIF512 and MIF1024
images.  
We therefore created an individual sky for each image by taking the median of
all the other images of the same exposure time while masking regions known
to include astrophysical objects in the first drizzled image.
We found that four MIF512 images were 
insufficient to make acceptable sky images;  however the sky
subtraction appeared to work well for the MIF1024 images. 
The  final combination of the eight sky
subtracted MIF1024 images shows no evidence of correlated noise along
the direction
of the dither pattern, and the noise  on arcsecond scales has been
reduced by more than a factor of two.
Furthermore, since the individual images are  read noise limited, the
neglected
MIF512 images represent only about one-eighth of the total signal-to-noise ratio
of the data, and therefore their loss does not significantly affect
the final signal-to-noise ratio that can be achieved.

The extended emission can be clearly seen in the NICMOS image displayed
in Figure~\ref{f:both}.  The same $1\farcs5$ aperture as used in the WFPC2 image
reveals a total count rate of $0.23 \pm 0.02$ counts per second in
the NIC2 F160W filter, which corresponds to $470 \pm 50$~nJy or a 
magnitude of $\H = 23.21 \pm 0.1$.    There is no sign of the OT in
the NICMOS image;  we find a $3 \sigma$ upper limit on the flux
density of the OT of $\H  \ge  25.9$ or $40$~nJy.

\subsection*{The Keck Images}

Images of the GRB 970228 field were obtained in the K-band on April 8, 1998
with the Keck I 10m telescope and NIRC \cite{ms94}.  A
9-point-pattern mosaic of individual 10-second exposures was obtained and
co-added to produce an image with total exposure time of 2160 seconds.  The
resulting image point spread function measured $0\farcs6$ FWHM.  The flux of
the extended source was measured inside a $1\farcs5$ 
diameter aperture centered on
the source, and an approximate aperture correction to $4''$ was made using the
curve of growth of the nearby star.  Calibration  was performed using the
infrared standards of Persson et al. (1998)\nocite{pmkr+98}, 
including an airmass
correction term.  We obtain $\K = 22.8 \pm 0.3$ or  $500 \pm 150$~nJy.
Thus, within the errors, the spectrum of the extended emission
is flat
in $f_{\nu}$ between H and K.

%
\begin{center}
\begin{tabular}{ccccc}
\multicolumn{5}{c}{{\bf Table 1:} Optical and Near-Infrared Photometry
of GRB~970228}\\
\hline
\hline
& & & \multicolumn{2}{c}{Johnson-Cousins Magnitudes}\\
Instrument & Band & Date (UT) & Optical Transient & Nebulosity \\
\hline
HST/WFPC2 & F606W/V & 26 Mar 1997 & $26.05 \pm 0.07^a$ & $25.8 \pm 0.3$ \\
HST/WFPC2 & F814W/I & 26 Mar 1997 & $24.10 \pm 0.07$ & $24.6 \pm 0.3$ \\
HST/WFPC2 & F606W/V & 07 Apr 1997 & $26.35 \pm 0.10$ & $25.7 \pm 0.3$ \\
HST/WFPC2 & F814W/I & 07 Apr 1997 & $24.65 \pm 0.10$ & $24.3 \pm 0.3$ \\
HST/STIS  & 50CCD/V & 04 Sep 1997 & $28.00 \pm 0.25$ & $25.8 \pm 0.25$ \\
HST/NICMOS2 & F160W/H & 24 Feb 1998 & $> 25.9^b$       & $23.2 \pm 0.1$ \\
Keck/NIRC   & K & 1998 Apr 08 &  --              & $22.8 \pm 0.3$ \\
\hline  
\multicolumn{5}{l}{$^a$ Errors represent 1-$\sigma$ uncertainties.}\\
\multicolumn{5}{l}{$^b$ 3-$\sigma$ upper limit.}\\
\end{tabular}
\end{center}

\subsection*{The NTT Images}
We have also re-examined the NTT observation of March 13 \cite{ggp+97} to
further test whether the nebular magnitude may have varied with time.
As in the case of our analysis of the STIS
image, we have again used the stellar image $\sim 2\farcs5$
to the east of the OT as a point spread function.    We find that
we can subtract a point source from the position of the OT  and
leave behind 
a  ``nebula''
as faint as, or fainter than, the extended emission
in the HST/STIS image without producing any noticeable
sky subtraction errors. 
Thus, we find no evidence
that the nebula has faded with time.

\section*{Discussion}

\subsection*{The Galactic Extinction in the Direction of GRB~970228}

GRB~970228 lies in the direction of $l = 188.9$ and $b = -17.9$.  As a
result, its
apparent visual magnitude and color are significantly
affected by interstellar extinction in our galaxy. Therefore, 
we have  used several techniques to estimate
the foreground
Galactic extinction in its direction, so that 
we can obtain the true colors and unextincted magnitudes
of the OT and the surrounding nebulosity.
The colors and counts of background galaxies suggest $A_V \sim 0.6 \pm 0.2$
\cite{gfd98}.
The catalogs of Burstein and Heiles (1982) and Schlegel \etal (1998) 
\nocite{sfd98} give
0.8 and 0.75 respectively, with errors of approximately 0.15 mag.  The color
of a background K4V star in the field
suggests $A_V = 0.9 \pm 0.2$, and the Na\,I absorption in its spectrum
implies a strong upper limit of $\sim 1.0$.  For the remainder of this
paper then, we will use the value $A_V = 0.8 \pm 0.2$, which is roughly
the average of the estimates.  However, because this value is noticeably
less than that published by Castander and Lamb (1997), we discuss our
derivation of the extinction in detail in an appendix.


\nocite{mbg+98}

\subsection*{The Time History and Spectral Behavior of the Optical Transient}

\begin{figure}[t!] 
\centerline{\epsfig{file=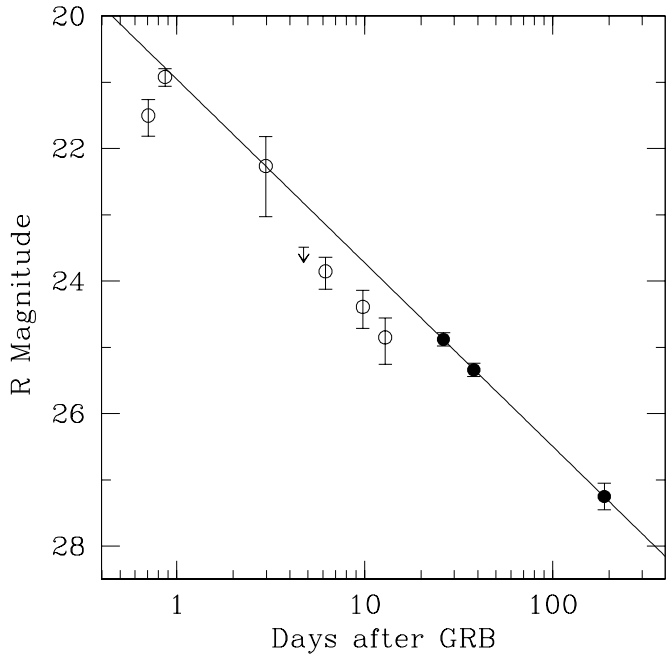,height=4.5in,width=4.5in}}
\vspace{10pt}
\caption{The R magnitude of the OT as a function of
time. Ground-based measurements are represented as open circles, and HST points as 
filled circles. 
A nebular R magnitude of 25.3 has been subtracted from all the
non-HST magnitudes, which have been  taken from Galama et al. (1998) and 
Masetti et al. (1998).  
An error corresponding to the uncertainty in
the nebular magnitude has been added in quadrature to the ground-based 
errors.
The line shows the best fit power law (index $-1.10 \pm 0.1$) through the three HST
observations.  Upper limits determined by ground based telescopes have been
reported only if falling below the power law curve.} 

\label{f:ot}
\end{figure}
In Figure~\ref{f:ot} we plot the magnitude of the OT as a function of time
since the burst on 28 February 1997.  The STIS magnitude has been converted
to R using the color determined from the WFPC2 V and I observations.
A power law
of the form $f(t) = a_0*t^{\alpha}$ has been fitted to the
HST points and extrapolated back to earlier times.  We find a best fit of
$\alpha = -1.10 \pm 0.1$.
A host R magnitude of 25.3 has been subtracted from all
non-HST magnitudes. 
This is the {\it minimum} host R 
brightness consistent with an interpolation of the V and I data.  If the
host is in fact somewhat brighter than this, a proper subtraction
would leave the first few points on the plot virtually unchanged, but
would somewhat {\it increase} the discrepancy between the
power-law extrapolation and the points taken 4 to
11 days after outburst.
The 
outlying points were taken at different observatories  and
reduced by different groups of observers.  One of these points is, in
fact, the NTT data that we have ourselves re-reduced as described
above.  We do not believe that these points can be discounted.

Keck observations of GRB~970228 taken on 30 and 31 March 1997 showed 
total
magnitudes, for the sum of OT and host, 
of $\K = 22.0$ and $\J=23.5$ \cite{snam+97}.   This K band magnitude
is about a factor of two greater than seen in our observations, one
year later.  Therefore,  it would appear that  in late March 1997, the
optical luminosity in K was equally divided between OT and extended
emission.
Similarly if we interpolate the J and K values to estimate an H
magnitude in late March 1997, and compare this with the observed HST F160W
magnitude, we again find that the power in the object is equally
divided between host and OT.  Furthermore, the estimated H and K flux densities
for the OT on March 30 are equal within the errors 
with an average value of $\sim 500$~nJy. 

Because the OT appears to have followed a  power-law decline between 26
March 1997 and 7 April 1997, we can use the power-law to interpolate
the F814W to the 31 March 1997.   Remarkably one finds that 
the F814W flux density is $500 \pm 50$~nJy.  Thus the spectrum
of the OT is flat in $f_{\nu}$ between 800 and 2200~nm.  In
contrast, even after correcting for foreground
Galactic extinction, the $\V - \I$ color of the OT implies
a spectrum $f_{\nu} \propto \nu^{-3}$.    Unfortunately, we do not
have infrared magnitudes for the OT at other times, and therefore
we cannot tell whether this break in the spectrum is constant with
time, or whether it results from strong temporal variability of the
OT in the infrared.  It is certainly too abrupt, however, to be due
to extinction in a host galaxy.    While strong theoretical arguments
exist for expecting temporal variability in the OT \cite{mrw98,rm98}, it is
far from clear that the variability that would be required (at least
a factor of 3 - 6 ) could occur in one region of the spectrum, while
another, an octave away, was unaffected.  However, a break in 
the spectrum at $1 \mu$m, or $3 \times 10^{14}$~Hz, is also not expected
at late times in the standard model 
(see, for example, Sari, Piran and Narayan 1998\nocite{spn98}), nor
is it seen in the well-studied (and well-behaved) spectrum of
GRB~970508 \cite{gwb+98}. 

Nonetheless, the fact that for over 180 days, the OT behaved roughly
as the predicted power-law \cite{wrm97} is quite remarkable -- for
as the fireball of the burst expands, it sweeps up the material
of the ISM surrounding it, and its power-law expansion should falter
when it has absorbed an ISM rest mass comparable to the energy of
the initial explosion, or when
\begin{equation}
t\approx 1\mbox{\,yr}\left(\frac{E_{52}}{n}\right)^{1/3},
\end{equation}
where $E_{52}$ is the initial energy of the explosion in
units of $10^{52}$ ergs and
$n$ is the density of the surrounding medium in protons
per cubic centimeter.   However, were
the GRB a Galactic rather than an extragalactic phenomenon, the amount
of energy available would only be of order $10^{41}$ ergs, and for
any imaginable density the break would occur on a timescale of days
rather than many months, although as noted by Panaitescu and M\'esz\'aros 
(1998)\nocite{pm98},
this break can be mitigated by a significantly non-spherical outflow
geometry.  
Nonetheless, the power law fit is in itself an 
argument for the extragalactic nature of the burst, and thus
is in agreement with
all the other evidence presented in this paper.   

\subsection*{The Nature of the Host}

If the extended emission under the OT is truly a host galaxy, rather
than, as has been proposed, a reflection nebula from an 
explosion in our own galaxy \cite{cl98},  
then one would 
expect it to have a size and color which is reasonable for  a galaxy
of its magnitude.  On  the other hand, if a large fraction of the sky is
covered by galaxies whose observed properties are similar to the proposed
host, then even if the nebulosity
is shown to be a galaxy, one would still have little evidence that it is indeed the host,
rather than a chance superposition of a galaxy along the line of sight.

\begin{figure}[H] 
\centerline{\epsfig{file=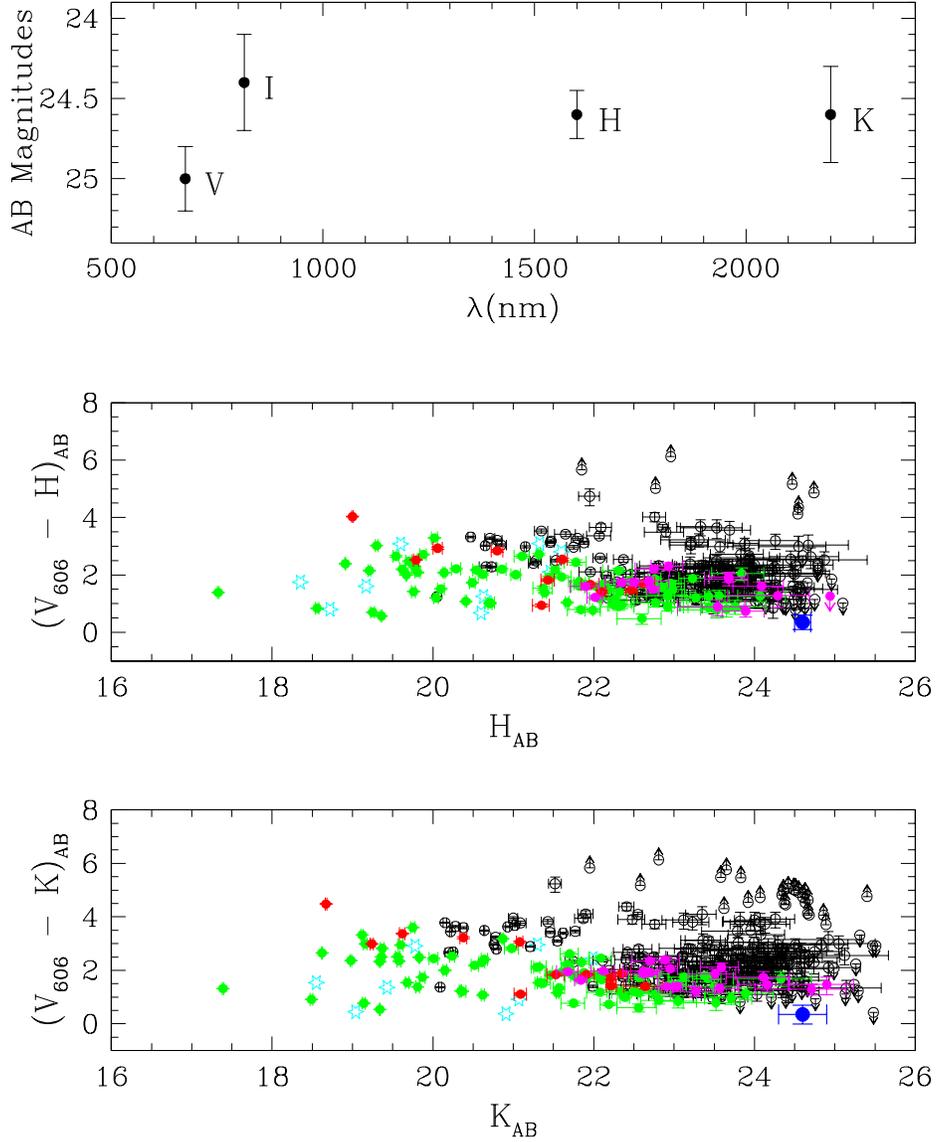,width=5.0in}}
\vspace{10pt}
\caption{In the upper panel, the extinction-corrected 
spectral energy distribution of the host galaxy is shown.
In the bottom two panels,  we place the host galaxy on a color-magnitude
diagram of objects in the HDF.  Stars are shown as light blue stars.
Galaxies in the HDF with spectrophotometric redshift are shown
in green for $0 < z < 1$, red for $1 < z < 2$ and magenta for
$z > 2$.  The host galaxy is shown in dark blue. 
Upper and lower limits are shown with arrows.}
\label{f:hdf}
\end{figure}

We have therefore compared the observed extended emission with galaxies
found in the HDF.
After adjusting for the differing depths and pixel sizes
of the STIS and HDF images and the extinction in the direction
of GRB~970228, we
find that only $\sim 2\%$ of the HDF is covered by galaxies
of comparable or greater surface brightness of the putative
host (25.3 mag/arcsec$^2$).  This result is
similar to, but tighter than, that of Sahu \etal 1997, and 
implies that the probability of a chance
superposition is negligible.  Furthermore,
we find that the area on the sky of the proposed host 
($\sim 0.5$~arcsec$^2$) is near the median
of galaxies of similar magnitude in the HDF.   The most remarkable result,
however, arises when we compare the color of the host with that of
other galaxies in the HDF.  In Figure~\ref{f:hdf}
we show the spectral energy
distribution of the host galaxy and compare it to the colors and
magnitudes of objects in the HDF.  
The $\H$ and $\K$ band HDF magnitudes are from KPNO/IRIM observations of
the HDF \cite{dick97}.  The F606W images of the HDF have been convolved
to the ground-based resolution and colors determined using
isophotal magnitudes.
We use AB magnitudes here rather than Vega magnitudes 
to conform to
other work that has been done on the 
HDF\footnote{
AB magnitudes are defined by the
relation ${\rm mag_{AB}} = -2.5 \log(f_{\nu}) + 23.9$, where $f_{\nu}$
is given in $\mu$Jy. The zeropoint of the AB magnitude system,
unlike the Vega magnitude system, does not depend upon wavelength;
however, the zeropoint of the AB magnitude system has been set 
so that it agrees with the Vega magnitude system
in V.}.

As can be seen in Figure~\ref{f:hdf}, the host of GRB~970228 is 
unusually blue.   
Indeed, it is as blue as any galaxy 
in the KPNO/IRIM HDF.   Although the completeness limit of this
sample is slightly brighter than the magnitude of the host,
HST/NICMOS
observations recently performed on the HDF should soon
allow us to test this relation on a sample which has a limiting
magnitude safely below the host magnitude.
Additionally, we find  that the host is too blue to be fit
by shifting 
local galaxy templates \cite{cww80} (including blue irregular
dwarfs, such as the Magellanic clouds) to higher redshift.
Therefore, to attempt to fit the host's colors, we created
a grid of synthetic stellar populations as described in Babul
and Ferguson (1996)\nocite{bf96}.  
The models used a Salpeter IMF from  $0.6$ to $100 \Msun$.  Metallicities
ranged from solar to 0.05 solar. The e-folding times of the
starburst were varied from $10^6$ to $10^{10}$ yrs, and the age at
time of observation from $10^6$ to $10^{10}$ yrs. 
The stellar populations were given internal extinctions 
from 0 to 2 magnitudes in $A_V$.
If we require only that a synthetic galaxy agree with the observed
colors to within the photometric errors ($\chi^2(3) < 4$),
we can create galaxies of nearly any age and
redshift which match.
However, if we also require that the synthetic galaxy lie
within the locus of observed galaxies seen in Figure~\ref{f:hdf}
(because of the photometric errors, a galaxy could match the 
host without lying inside the locus), the set of possible
populations is dramatically reduced.  All
stellar populations older than $2 \times 10^8$ yrs which match 
have essentially constant star formation rates (the time
of observation is much smaller than the e-folding time of the burst);
furthermore, 
these model galaxies cluster around two redshifts, $z \sim 0.8$ and
$z \sim 4.2$.  In the former case, the model is placing the 4000~\AA\
break between F606W and F814W, and in the latter case the F606W flux
is suppressed by the Ly$\alpha$ forest.   Interestingly, at a
redshift of $z \sim 0.8$ the magnitude of the host would imply
an intrinsic luminosity comparable to that of the large Magellanic cloud,
and based on its photometry (and a conversion of UV luminosity to star 
formation rate, SFR, (Kennicutt 1998)  \nocite{ken98} would have have an SFR of $\sim 0.5 \Msun \,{\rm yr}^-1$, assuming ${\rm H_0} = 65$ km/s and $\Omega = 0.3$;
however,  at $z \sim 4$, it would be unusually luminous -- 
noticeably brighter than any of the
photometrically identified $z \sim 4$ candidates in the HDF \cite{mdg+96},
and have an SFR of $\sim 13  \Msun \,{\rm yr}^{-1}$.

One can perhaps further limit possible host galaxies using an absence
of evidence:  in spite of the expenditure of many hours of spectroscopic
time on Keck, this object apparently displays no emission lines.  
However, a young starburst -- or even a continuous starburst -- should
show [O\,{\sc ii}] (3727 \AA) emission \cite{ken92} and in many cases would
show Ly$\alpha$ (1216 \AA) \cite{sgpd+96}, 
though this line can be suppressed by
resonant scattering and absorption.   For those lines to be present,
yet unobserved, the redshift of the host would need to be in the
range $1.3 < z < 2.5$ -- where O[II] has left the optical spectroscopic
window in the red, yet Ly$\alpha$ has not yet entered in the blue.
Thus the continuous star formation models mentioned above
are probably ruled out.  The reason for the extraordinary blue color
of the host is most likely the obvious reason -- it is forming stars
at an extraordinary rate for its mass.  Indeed, all of the preferred stellar
population models predict that the doubling time of the stellar
mass is $\simlt 2 \times 10^8$ yrs, and it is this doubling time,
rather than the star formation rate of the host, which will best
indicate whether the photometry suggests that GRBs are linked with
star formation.  Indeed, from the slope of faint galaxy counts
it can be inferred that in the range $1 \simlt z \simlt 3$ the
luminosity function of galaxies must have been far steeper than
is found locally \cite{hogg98,hf98}.  Thus the majority of star formation
at these redshifts may be found in galaxies with luminosities
(and SFRs) well below $L_*$, the knee of the
luminosity function \cite{she76}.  Therefore, if GRBs are
associated with star formation, it will be the colors (and
stellar lines) rather than the absolute luminosities of the
hosts which will provide the proof.

Additionally, the comparison of the extinction-corrected color of the host with
the HDF  provides a good indication that we have not underestimated
the foreground
Galactic extinction.  A larger $\AV$ than the
$0.75$ used would have made the object bluer, and caused it to lie
even further outside the locus of objects seen in the HDF.  Indeed,
comparison with the HDF suggests that the somewhat lower extinction
of $A_V = 0.6$ implied by the comparison of the colors of background
galaxies may have been closest to the correct extinction.  However,
even if we were to reduce the assumed extinction to match
this estimate, the host would remain unusually blue.

\subsection*{Star Formation and GRBs}

Although the host of GRB~970228 may be the first host galaxy to have sufficient
published multicolor imaging to allow one to constrain the 
SFR, it is not the first host with strong evidence of star-formation.
All three GRBs with measured cosmological redshifts lie in hosts
which display prominent emission from lines associated with star-formation, and
in all three cases the strength of those lines is high
for galaxies of comparable magnitude and redshift \cite{dkg+98,kdrg+98,bkdf98a}.
Although other host galaxies have been observed spectroscopically (Kulkarni,
Metzger personal communications),  the fact that no other hosts
show lines is not necessarily an argument against strong star
formation in these galaxies.  As noted by 
Hogg and Fruchter (1998)\nocite{hf98} 
reasonable models of the cosmological density of GRBs with redshift
predict a substantial fraction in the range of $1.3 < z < 2.5$ where,
as mentioned earlier, no strong lines are found in the optical spectroscopic
window. 
Thus, one can argue that in all cases where 
the observations allow one to determine the
SFR, that rate has been found to be high. 

Unfortunately, our data does not allow us
to distinguish between the two
leading candidates for the creation of bursts, binary neutron star
\cite{npp92} and hypernovae \cite{bp98}, as binary-neutron star
coalescence should occur on a timescale comparable to that
of many star bursts ($\sim 10^8$~yrs) \cite{lsp+98}.  
The primary difference one expects for these two models is that
due to kicks imparted to the neutron stars at birth, neutron star
binaries should occasionally be well outside of a galaxy when the
burst occurs \cite{bsp98}.  However, at present, we are only able
to localize bursts using X-ray, radio or optical
afterglows.  Yet the strength of the afterglow
may depend critically on the density of the surrounding medium \cite{mr97}. 
The low density medium
of intergalactic space may be too
tenuous to produce detectable afterglows.  In this case,
deciding between binary neutron stars and hypernovae as
the progenitors of GRBs may require  a new generation of
space observatories capable of localizing the burst on the
basis of gamma-rays alone.

\section{Appendix:  The Foreground Galactic Extinction towards GRB~970228}

In this appendix we present several different independent methods that
we have used to estimate the extinction in the direction of GRB~970228.
We have consulted
the extinction catalogs of Burstein and Heiles (1982) \nocite{bh82} and 
Schlegel, Finkbeiner and Davis (1997), and
have attempted to directly determine the
extinction from the counts and colors of
galaxies in the WFPC field, as well as from
spectroscopy and imaging of Galactic stars in
the field.  All these methods give consistent
results.  

\subsubsection*{All Sky Extinction Maps}

Burstein and Heiles  (1982) have used 
measured HI column densities and an
empirically determined constant of conversion between HI to
extinction, in combination with background
galaxy counts to derive a map of Galactic
extinction with a spatial resolution of about
$0\fdg6$.  Based on their work, the extinction
in the direction of GRB~970228 is $A_V = 0.8 \pm 0.1$.
Similarly, Schlegel \etal (1997)\nocite{sfd97} 
have used the IRAS
$100 \mu$ maps to estimate foreground Galactic
extinction, with an angular resolution of $\sim 6'$.
Their work implies an extinction of $A_V = 0.75 \pm 0.15$.
The consistency of these numbers   is encouraging,
and as will be seen agrees well with our other estimates. 

\subsubsection*{Galaxy Counts and Colors}

Foreground extinction should be visible as a diminution
in the number counts of galaxies in the field to a given
limiting magnitude, and a reddening
of those which are detected.
We therefore have compared the WFPC2 images of the GRB~970228
field with those of four high latitude fields taken from the HST
archive, the HDF \cite{wms96},
the field around the weak radio galaxy 53W002 \cite{pwk+96},
and  the fields at $\alpha_{2000}=15^{{\rm h}}58^{{\rm m}}49\fs8, 
\delta_{2000}=42^{\circ}05^{\prime}23^{\prime\prime}$,
and at $\alpha_{2000}=14^{{\rm h}}17^{{\rm m}}43\fs63, \delta_{2000} =
52^{\circ}28\arcmin41\farcs2$
imaged by Westphal and collaborators as part of HST proposal 5109.
In the case of the HDF, the distributed drizzled images were used.
In all other cases we reduced the data ourselves, again producing
a drizzled output.  Further details of the data reduction technique can
be found in Gonz\'alez, Fruchter and Dirsch
(1998)\nocite{gfd98}.

The SExtractor object extraction software \cite{ba96}
was used to locate objects,  perform star-galaxy separation
and obtain photometry on galaxies.   Objects were extracted down
to a limiting surface brightness in the F606W images
of $25\ {\rm mag}$ arcsec$^{-2}$, which was determined
by the shallowest
image  (that of the GRB field).
Colors
were obtained by placing matching fixed-size apertures of
diameter $0\farcs4$ on the F814W fields.
To simulate extinction the limiting magnitude was progressively
raised in the reference fields until the number counts and
colors matched those of the GRB field.  A more detailed
account of the actual fitting procedure can be found in
Gonz\'alez \etal (1998). 

Comparison of the counts in the GRB field with those of the
HDF and 53W002 field gives a best estimate of extinction
of $A_V \sim 0.7 \pm 0.2$; however, 
the two Westphal fields have significantly
fewer galaxies, by an amount that is large compared to the
expected clustering error (though, as discussed in
Gonz\`alez \etal 1998, this error may be generally underestimated).
If these two
fields are averaged in with the HDF and 53W002 fields,
the best value of extinction
is  $A_V = 0.5 \pm 0.2$.  
Although the counts differ, the colors of background galaxies
in these four fields agree well.    The extinction
implied by the average color difference between the GRB field
and the four background fields is $A_V = 0.6 \pm 0.1$,  
where the error is 
the measured dispersion
in estimated extinction between the four reference fields using
the color index.     
However, there may
be a slight bias in the color estimation -- if the reddening
is clumpy, the most reddened 
background galaxies may fall out of the sample and not affect
the average color.   Thus the estimate of $A_V = 0.6$ should probably
be viewed as a lower limit.  In any case, this estimate is
in good agreement with the Burstein
and Heiles (1982) and Schlegel \etal (1998) extinctions.

\subsubsection*{Stellar Spectroscopy}

We have also examined the long-slit spectra of the GRB~970228 field obtained by
Tonry \etal (1997a,b)\nocite{thc+97a,thc+97b}, and generously made 
available
on the World Wide Web,  to estimate the
extinction to GRB~970228.  
The observations were done using the LRIS spectrograph mounted on the Keck
II telescope.  The final combined spectrum covers the
wavelength range 4320\,$-$\,9300 \AA\ at a spectral resolution of 11 \AA\
(FWHM). The total exposure time was 5000 s, and the weighted mean
airmass was 1.52. The slit was 1$''$ wide.   We performed flux
calibration
using a spectrum of spectrophotometric standard star Hiltner
600 \cite{hsh+94} with the same slit width.
Three stars are present in the slit of the  Tonry \etal spectrum,
denoted S1, S2 and S3 (see Tonry \etal 1997b). Before extracting the
spectra of these stars, we corrected the summed long-slit spectrum for
tilt and geometric distortion by measuring the centroids of the stars
along the slit as a function of wavelength. The table of centroid
positions was then used to rectify the long-slit spectra, using IRAF
tasks {\it geomap\/} and {\it geotran}. The star spectra were
subsequently extracted using a ``optimal extraction'' algorithm,
described by Horne (1986)\nocite{hor86}.

The slit in these observations falls
along two K stars in the PC field of the WFPC2 images (denoted
as S1 and S2 as in Tonry \etal 1997b), as well
as on the GRB itself.  The brightest of these two stars can be
clearly identified as a K4V, and it is likely that the other
is also a K4V.  Like Castander and Lamb (1998), we have attempted
to fit the observed Keck continuum by applying varying extinctions
to library K4V spectra \cite{jhc84,sc92}. 
{\it We find no satisfactory solution}.  
We cannot match the slope across the entire continuum no matter
what extinction we use (examples of  poor fits can be seen in
Figure 3 of Castander and Lamb).  This may be due partially
to the fact that the position angle of the slit was approximately
$45^{\circ}$ from the parallactic angle.  However, this cannot
entirely account for the effect.  It appears possible that the stars
were partially off the slit  and/or that the publicly available flat
field does not provide a true spectrophotometric correction.

While we may not be able to properly flatten the Tonry \etal (1997b) spectra
on large scales,
this should not significantly affect our ability to examine the equivalent
widths of lines in the spectrum.   In particular, we can use the
equivalent width of the  interstellar Na\,I
$D$ doublet ($\lambda\lambda\,5890, 5896$ \AA) to place a limit
on the reddening in the field.

In the interstellar medium (ISM), gas and dust grains are generally
associated with one another. This was recognized in the 1940's
from the correlations between the intensity of the interstellar Na\,I
$D$ doublet and the reddening of
starlight (e.g., Spitzer 1948)\nocite{spi48}.
Indeed, observations of stars at
intermediate and high Galactic latitudes indicate that the column
density of interstellar Na\,I is correlated with \EBV\  
\cite{coh74,hob74a,fvg85}.
This relationship shows
however a rather significant scatter, which can be understood as
being due to {\it spatial\/} variations: for  
disk stars behind or within dense dark clouds in the solar neighborhood,
the equivalent width (EW) of interstellar Na\,I in spectra of stars
is smaller (for a given amount of reddening) than those of more
distant stars at high Galactic latitude. This is due to Na\,I
depletion in dense clouds, where many heavy elements must be locked
onto dust grains (e.g., Cohen 1973\nocite{coh73}). Another contributor to the
scatter in the $EW$(Na\,I) vs.\ \EBV\ relationship (at high column
densities) is the effect of ionization:\ as the cloud density increases
the ratio of neutral to ionized gas increases proportionally, causing
{\it enhanced\/} Na\,I absorption for a given \EBV\ \cite{hob74a}.

\begin{figure}
\centerline{
\psfig{figure=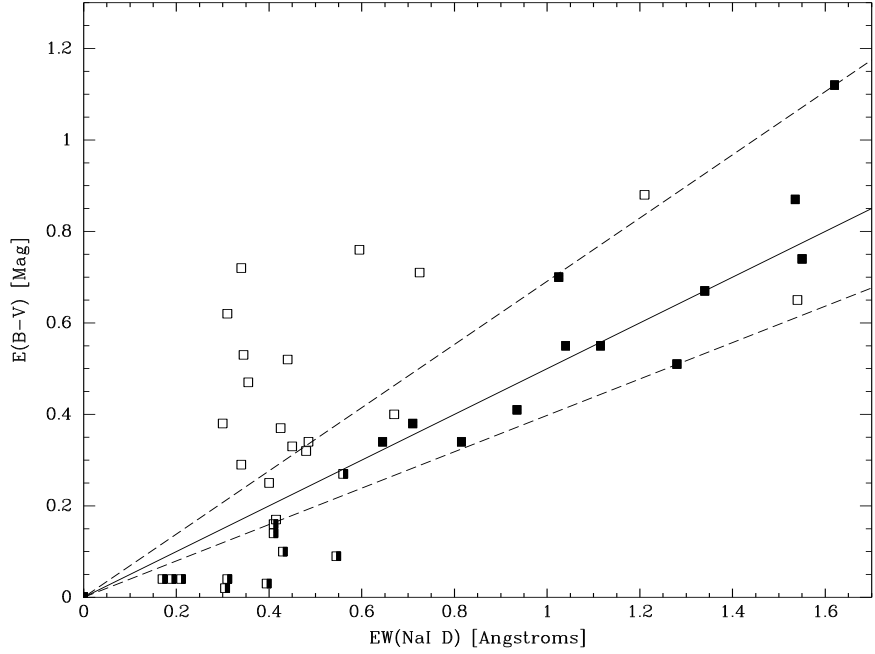,angle=-90.,width=4.5 in}
}
\caption[]{Equivalent width of Na\,I $D$ absorption line against \EBV\
for stars along different lines of sight. Open symbols are for nearby
stars within or behind dense clouds in the solar neighborhood,
half-open symbols are for stars at medium to high Galactic latitude,
and solid symbols are for distant supergiant stars for which the line
of sight goes through the low-density intercloud medium. The solid
line is a linear least-square fit to the data on the latter category
of stars, and the dashed lines delineate the extreme possible slopes of that
relation. See text for discussion.} 
\label{f:EW_vs_EBV}
\end{figure}

In order to quantify the interstellar Na\,I absorption vs.\ reddening
relationship, we show in Fig.\ \ref{f:EW_vs_EBV} $EW$(Na\,I) against \EBV\
for a large number of early-type (O and early B) stars taken from the
literature \cite{coh73,coh74,coh75,hob74b,hob78,fvg85}.
So that we could compare these data to the Tonry et al.\
spectra, which were obtained with a lower dispersion, we selected stars for
which both components of the Na\,I doublet were measured. 
Fig.\ \ref{f:EW_vs_EBV} depicts these data for different lines of
sight: {\it (i)\/} nearby stars within or behind dense clouds, {\it (ii)\/}
distant supergiant stars for which the line of sight goes through the
low-density intercloud medium, and {\it (iii)\/} stars at high
Galactic latitude, for which the reddening arises most probably from
the disk layer with a half-thickness of 100 pc \cite{coh74}. 
Case {\it (ii)\/} is
arguably the most appropriate for the line of sight to GRB~970228,
since Galactic globular clusters at low-latitude show a
similar relation between $EW$(Na\,I) and
\EBV\ \cite{ba86}. 

In K main-sequence stars such as S1 and
S2, the Na\,I $D$ absorption line strength is partly due to the
stellar photosphere.  Furthermore, contribution of the stellar
photosphere to the total line strength is a function of stellar
metallicity. To quantify the stellar contribution, we have measured the
equivalent widths ($EW$) of the Na\,I $D$ doublet as well as the Mg\,I $b$
triplet (which is a measure of stellar metallicity, and unaffected by
the ISM) for stars S1 and S2 and compared them with those of the
library K4V star from Silva and Cornell (1992)\nocite{sc92}. 
The wavelength intervals for the line
indices were taken from Burstein \etal (1984)\nocite{bfg+84}. 
The results are
listed in Table~2.

\begin{table}
\begin{center}
\begin{tabular}{lcc}
\multicolumn{3}{c}{{\bf Table 2:} Na\,I and Mg\,I Equivalent Widths (in 
\AA)}\\
\hline
\hline
Star & $EW$(Na\,I $D$) & $EW$(Mg\,I\,$b$) \\
\hline
S1 & $6.1\pm1.0$ & $9.0\pm1.0$ \\
S2 & $6.2\pm0.2$ & $8.4\pm0.2$ \\
K4V$^a$ & $6.0\pm0.1$ & $7.1\pm0.1$ \\
\hline
\multicolumn{3}{l}{$^a$ Library star from Silva \& Cornell (1992).}\\
\end{tabular}
\end{center}
\end{table}

It is evident from Table~2 that stars S1 and S2 are more
metal rich than the library K4V star, and yet have Na\,I $D$
equivalent widths that are at most marginally larger. The
straightforward interpretation of this result is therefore that the
interstellar contribution to $EW$(Na\,I $D$) (hereafter $EW$(Na\,I
$D$)$_{\scrm{ISM}}$) for stars S1 and S2 is consistent with
zero. 
To derive a stringent upper limit to $EW$(Na\,I
$D$)$_{\scrm{ISM}}$, we used the Na\,I $D$ profile of the K4V library
star as template (i.e., purely stellar) spectrum and evaluated the maximum
``extra'' equivalent width that can be accommodated in the spectrum of
star S2 within the errors (the spectrum of star S1 is too noisy 
for us to use for this purpose). The resulting upper limit is $EW$(Na\,I
$D$)$_{\scrm{ISM}}$\,$\la$\,0.5 \AA. 

Using the relation of \EBV\ vs.\ $EW$(Na\,I $D$)$_{\scrm{ISM}}$ for distant
stars in the Galactic disk (depicted in Fig.~\ref{f:EW_vs_EBV} as a
solid line, along with two dashed lines that depict the minimum and
maximum possible slopes), we find that this upper limit to $EW$(Na\,I
$D$)$_{\scrm{ISM}}$ translates into \EBV\ $\la$ 0.25$^{+0.09}_{-0.05}$
\AA, equivalent to  $A_V \la 0.75^{+0.27}_{-0.15}$ assuming the
Galactic extinction law of Rieke and Lebofsky (1985)\nocite{rl85}. 

\subsubsection*{Stellar Colors}
There is one further means of measuring the extinction available to
us.  Given that we believe star S2 of Tonry \etal (1997b) can be
accurately classified as a K4V, we can examine the F606W and F814W images
and compare the observed colors with those of library stars.
We have done this using the IRAF/STSDAS \cite{tody93} program
{\tt synphot} to multiply a standard library spectrum (chosen
from the libraries mentioned above) with
an extinction curve and
the response of the WFPC2 instrument.
We find that the ratio of counts per second between the F606W and
F814W images of 1.12 is fit with an $A_V = 0.9 \pm 0.2$, where
the error is dominated by the choice of library spectrum and form
of the Galactic extinction curve.   Using a reddened library
spectrum in {\tt synphot} we find that S2 has an R band magnitude of 22.85.
After dereddening this would correspond, for a K4V star, to a distance
of $\sim 6.8$~kpc.  Given that this star is in the Galactic anticenter, and
at a Galactic latitude of $\sim 18^\circ$, all of the dust along
this line of sight is almost certainly between us and the star. 
Therefore, the reddening of the star is a valid measure of the
total extinction in this direction of the sky.

\section*{Acknowledgements}

We thank the Director of STScI, Bob Williams,
for allocating Director's Discretionary time to
observe GRB~970228 using STIS.  We benefitted
from discussions on the nature and redshift
distribution of hosts with David Hogg, and
on the magnitude of the foreground Galactic
extinction with Francisco Castander and Don Lamb.
We are also grateful to Mark Dickinson for providing us with a catalog
of the HDF/IRIM detections.


\end{document}